# Interfacial Charge Transfer Circumventing Momentum Mismatch at 2D van der Waals Heterojunctions

Haiming Zhu[1,2,*], Jue Wang[1,*], Zizhou Gong[3], Young Duck Kim[4], Martin Gustafsson[1], James Hone[4], Xiaoyang Zhu[1,†]

[1.] Department of Chemistry, Columbia University, New York, NY 10027, USA

[2.] Department of Chemistry, Zhejiang University, Hanzhou, Zhejiang, 310027, China

[3.] Department of Physics, Columbia University, New York, NY 10027, USA

[4.] Department of Mechanical Engineering, Columbia University, New York, NY 10027, USA

ABSTRACT:

Interfacial charge separation and recombination at heterojunctions of monolayer transition metal dichalcogenides (TMDCs) are of interest to two dimensional optoelectronic technologies. These processes can involve large changes in parallel momentum vector due to the confinement of electrons and holes to the K-valleys in each layer. Since these high-momentum valleys are usually not aligned across the interface of two TMDC monolayers, how parallel momentum is conserved in the charge separation or recombination process becomes a key question. Here we probe this question using the model system of a type-II heterojunction formed by $MoS_2$ and $WSe_2$ monolayers and the experimental technical of femtosecond pump-probe spectroscopy. Upon photo-excitation specifically of $WSe_2$ at the heterojunction, we observe ultrafast (<40 fs) electron transfer from $WSe_2$ to $MoS_2$, independent of the angular alignment and, thus, momentum mismatch between the two TMDCs. The resulting interlayer charge transfer exciton decays via nonradiatively recombination, with rates varying by up to three-orders of magnitude from sample to sample, but with no correlation with inter-layer angular alignment. We suggest that the initial interfacial charge separation and the subsequent interfacial charge recombination processes circumvent momentum mismatch via excess electronic energy and via defect-mediated recombination, respectively.

---

[*] These authors contributed equally.
[†] To whom correspondence should be addressed. E-mail: xyzhu@columbia.edu

Graphene-like two dimensional (2D) materials offer a unique platform for creating Van der Waals heterojunctions with emerging physical properties and unprecedented versatility from a growing library of 2D monolayer building blocks.[1–3] Among them, heterojunctions formed by stacking different transition metal dichalcogenide (TMDC) monolayers with type II band alignment have been particularly interesting for optoelectronic,[4,5] photovoltaic,[6–8] and spin-valleytronic[9,10] applications, because of the direct bandgap nature,[11,12] spin/valley polarization[10,13] and strong light-matter interaction[14] of constituent TMDC monolayers. In a type II heterojunction, exciton created by photo-excitation in one layer can dissociate by rapid electron or hole transfer across the interface (Fig. 1A). Because of reduced dielectronic screening in 2D, a spatially separated electron-hole pair across the interface forms interlayer charge transfer (CT) exciton with binding energy > 100 meV.[15,16] Previous studies have shown ultrafast (≤ 100 fs) interfacial CT processes in several 2D heterojunctions with arbitrarily aligned crystallographic orientations,[17,18] but radiative recombination[19,20] likely only occurs for small angular mismatches (<5°) as defined by the so-called light cones.[9,21] Because the conduction band minimum (CBM) and valence band maximum (VBM) in TMDCs are located at K points (or valleys) in momentum space with large parallel momentum vectors, the interfacial charge transfer process can be accompanied by large momentum change, Fig. 1B. Similar to inter-band scattering or electron-hole recombination in indirect bandgap bulk semiconductors,[22] the rates of charge separation and recombination across a TMDC hetero-interface should strongly depend on the magnitude of momentum change and phonon population (temperature). Since the angular alignment at 2D Van der Waals heterojunctions is quantitatively controllable,[23–25] a central question is whether or how the crystallographic alignment, thus the momentum mismatch, affect interfacial charge separation and recombination in 2D heterojunctions.

Here we experimentally examine the influence of interlayer twist angle (Δϕ) on the interfacial charge transfer dynamics at $WSe_2/MoS_2$ heterojunctions. $WSe_2/MoS_2$ forms a type II heterojunction with conduction band offset of 0.76 eV (CBM of the composite interface in $MoS_2$) and valence band offset of 0.83 eV (VBM in $WSe_2$).[26] We prepared $WSe_2/MoS_2$ heterojunctions with Δϕ varying in the complete range of 0-30 degrees using mechanical exfoliation and Van der Waals stacking.[27] We probe photo-induced electron transfer from $WSe_2$ to $MoS_2$ and subsequent interlayer charge recombination using optical pump-probe spectroscopy. We observe ultrafast (< 40 fs) interfacial electron transfer from $WSe_2$ monolayer to $MoS_2$ monolayer, independent of twist angle $\Delta\phi$. The interlayer charge recombination occurs on a much longer timescale and varies significantly from sample to sample (from ~ 40 ps to ~ 3ns), but showing no clear correlation with twist angle $\Delta\phi$. Exciton/carrier density dependent studies rule out the Auger scattering for interlayer excitons with excitons or doped carriers. These results can be well explained by the initial ultrafast resonant charge transfer across the interface with excess energy after photoexcitation and subsequent defect assisted nonradiative electron-hole recombination across the interface.

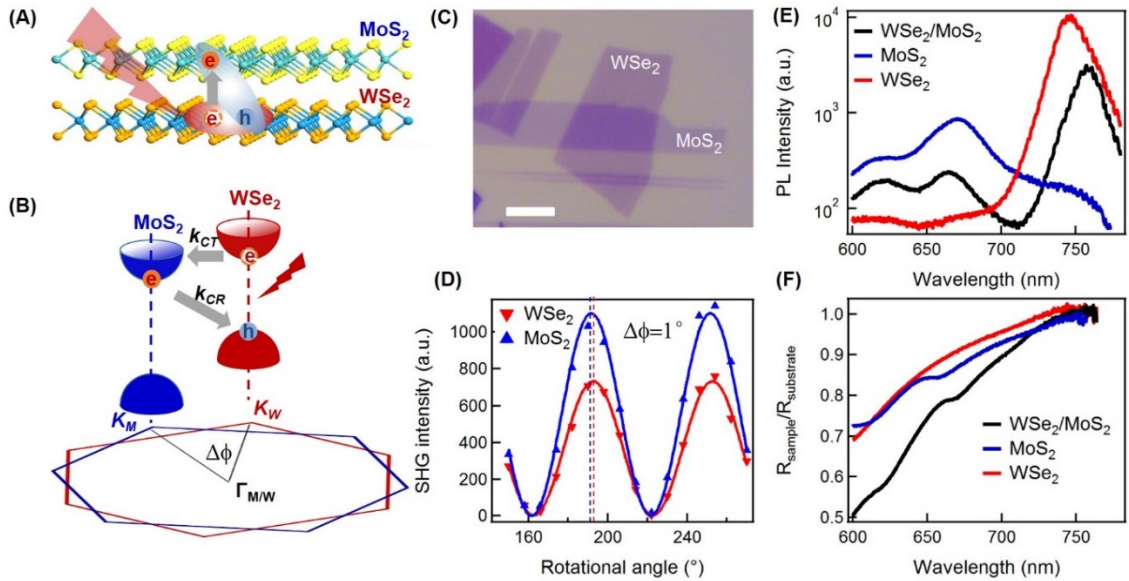

Fig. 1. (A-B) Schematic illustration of $WSe_2/MoS_2$ heterojunction where photoexcitation creates intralayer exciton in $WSe_2$ followed by electron transfer to $MoS_2$, forming indirect CT exciton in real space (A) and in momentum space (B). $\Delta\phi$ (between 0 and 30°) is the twist angle of relative crystallographic orientations or momentum space orientations for two monolayers. (C) Optical image of stacked $WSe_2/MoS_2$ heterojunction. Scale bar: 5 μm. (D) Second harmonic generation (SHG) intensity from $WSe_2$ and $MoS_2$ monolayer regions as a function of crystal's rotational angle after stacking. The rotational angle starts from arbitrary value. The solid lines are fits to the analysis described in **Methods**. The fits allow us to $\Delta\phi$ values to ±1°. (E-F) Photoluminescence (E) and reflectance (F) measurements of the $WSe_2$ monolayer, the $MoS_2$ monolayer and the $WSe_2/MoS_2$ heterojunction regions, respectively. Note the logarithm scale for PL intensity in panel (E).

We prepared $WSe_2/MoS_2$ heterojunctions by mechanical exfoliation and the polydimethylsiloxane (PDMS) transfer technique[28] (for the gated sample only) or the hexagonal boron nitride (h-BN) pickup/transfer technique[29] for all other samples (see **Methods**). Fig 1C shows optical image of a WSe$_2$/MoS$_2$ heterojunction on the SiO$_2$/Si substrate. This architecture allows us to access and characterize both the monolayer regions and heterojunction region. To determine the twist angle Δϕ between WSe$_2$ and MoS$_2$, we measured the intensity of linear-polarized second harmonic generation (SHG) as a function of crystal rotational angle ϕ,[30] Fig. 1D. We fit the angular dependent SHG intensity with six-fold symmetry (see **Methods** for details). Similar optical microscope images, as well as SHG and photoluminescence (PL) characterizations for all other samples are shown in Supporting Information (Fig. S1). For all the heterostructure samples used, the interlayer twist angle Δϕ varies between 0 and 30 degrees. Unlike the valance band edges with large spin-orbital (SO) splitting (~150 meV and ~450 meV for molybdenum[31,32] and tungsten dichalcogenides[33], respectively) and spin-valley coupling, the conduction band edges with opposite spins in MoS$_2$ or WSe$_2$ monolayer are nearly degenerate (SO splitting of 4 meV for MoS$_2$ conduction band edges[32]). Therefore, we can assume six-fold symmetry for electron in K and K' valley conduction band edges and Δϕ = 0 (30°) corresponds to the smallest (largest) momentum change for electron transfer from the WSe$_2$ K valley to the MoS$_2$ K valley. The same is true for subsequent recombination of an electron in MoS$_2$ with the hole left in WSe$_2$ at longer timescales. Based on the lattice parameters of MoS$_2$ (3.2 Å) and WSe$_2$ (3.3 Å), we obtain the momentum change from the K valley in one layer to the K valley in another as $\Delta K_\parallel$ = 0.07 Å$^{-1}$ and 1.25 Å$^{-1}$ for Δϕ = 0 and 30°, respectively.

We characterized the as-prepared WSe$_2$/MoS$_2$ heterojunctions with PL and reflection spectroscopy. The PL and reflection spectra ($R=R_{sample}/R_{substrate}$) of monolayer regions and heterojunction regions are shown in Figure 1E and 1F, respectively. The reflection spectrum of MoS$_2$ monolayer shows two clear dips at ~ 658 nm and ~ 610 nm, corresponding to the A and B excitonic resonances, respectively.[11] The WSe$_2$ monolayer does not show any distinct reflection peak (or dip) but a broad feature at ~ 745 nm, corresponding to the WSe$_2$ A exciton transition.[34] The opposite contribution of MoS$_2$ and WSe$_2$ excitonic resonance in the reflection spectrum results from an interference effect from the substrate SiO$_2$ layer; this will also affect the sign of transient reflection signal discussed later. These excitonic features in the reflection spectrum are slightly redshifted and broadened in the heterojunction region. The PL spectra of the monolayer regions show emission peaks from the A and the B excitons in MoS$_2$ and the A exciton in WSe$_2$. These emissions are significantly quenched in the heterojunction region (note the logarithmic intensity scale), suggesting efficient interfacial CT processes. A careful examination of PL spectrum in the near-IR range (800 to 1700 nm) shows no interlayer CT exciton emission, suggesting the dominance of nonradiative recombination of the interlayer electron-hole pair in our samples. This issue will be discussed in detail later.

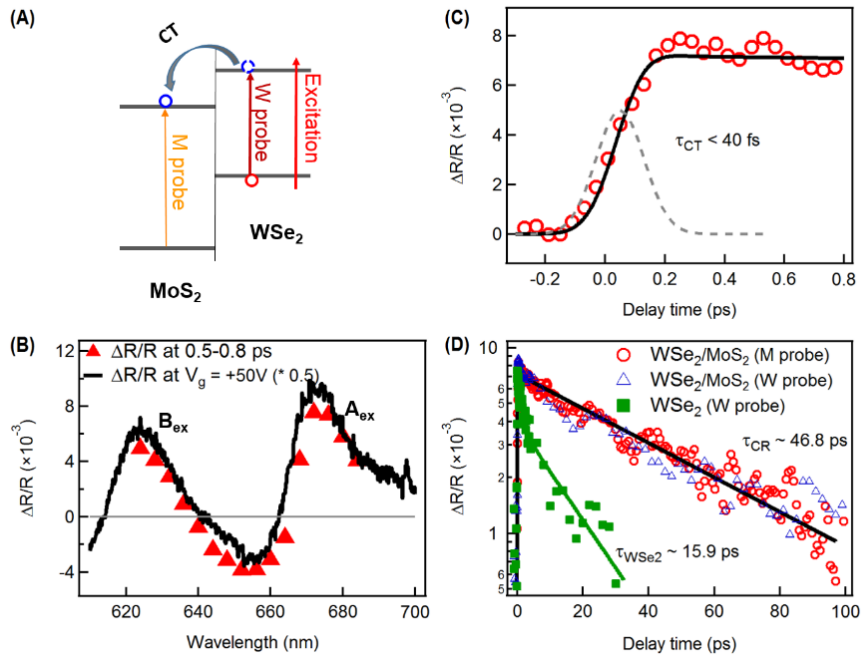

Fig. 2. (A) Scheme illustration of pump-probe measurement. We exclusively excite the A exciton in $WSe_2$ and probe $MoS_2$ (M probe) and $WSe_2$ (W probe) excitonic transitions, respectively. (B) Transient reflection spectrum (red triangle) after electron transfer (delay time 0.5~0.8 ps) and its comparison with reflection spectrum change of the heterojunction gate-doping at $V_g = +50V$ (black curve), confirming photoexcited electron transfer from $WSe_2$ to $MoS_2$. (C) Early time (< 1 ps) transient reflection kinetics from probing the $MoS_2$ transition with excitation of the $WSe_2$ A exciton, showing ultrafast electron transfer rate (< 40 fs). Gray dashed line is the pump-probe cross correlation function with FWHM ~ 190 fs. (D) Comparison of heterojunction kinetics under M probe and W probe condition at later times (up to 100 ps), showing interlayer electron-hole recombination process at the heterojunction. Also shown is the transient reflection kinetics of $WSe_2$ monolayer with a much shorter lifetime. Note the logarithm scale for vertical axis.

We probe interfacial charge separation and subsequently charge recombination processes at the $WSe_2/MoS_2$ heterojunction using pump–probe transient reflectance spectroscopy. We first excite the sample using a fs pulse and after a certain delay time, measure the reflection change (ΔR/R) of a broadband probe pulse (see **Methods**). Herein, we use 710 nm light to only excite the lowest energy A exciton in the $WSe_2$ monolayer to initiate the electron transfer from $WSe_2$ to $MoS_2$. We first investigate the exciton dynamics in $WSe_2$ monolayer region by probing the $WSe_2$ A exciton transition (W probe). The transient reflection spectrum of $WSe_2$ monolayer after photoexcitation is shown in Supporting Information, Fig. S2. It shows a negative transient reflection peak at ~ 741nm, corresponding to the bleach of A exciton resonance in $WSe_2$ monolayer due to the band filling of photoexcited excitons. The transient reflection signal rises promptly within experimental time resolution and its decay can be fitted by a biexponential function consisting of a 1.5 ps component (40%) and a 16.8 ps component (60%) with an average exciton lifetime of ~ 16 ps (green square/line in Fig. 2D). This lifetime is similar to the values reported previously[35,36] and is significantly shorter than the exciton radiative lifetime at room temperature (nanosecond).[37,38] The much shorter lifetime for the $WSe_2$ monolayer probed here can be ascribed to defect mediated nonradiative e-h recombination.[35]

Fig. 2B shows a transient reflection spectrum in the $MoS_2$ transition range (M probe, red triangles) obtained at 0.5~0.8 ps after photoexcitation of $WSe_2$ in the heterojunction. The positive peak at ~675 nm corresponds to bleach of the $MoS_2$ A exciton transition (the excitonic resonance shows a dip in reflection spectrum, Fig. 1F). In control experiments, we observe negligible signal in this wavelength range from the $MoS_2$ or $WSe_2$ monolayer region following photoexcitation at the same wavelength (710 nm). As comparison, we also show a difference reflectance spectrum, black curve in Fig. 2B, from the n-type gate-doping of $MoS_2$ in the heterojunction. Here the difference reflectance spectrum is obtained from subtracting $R_{sample}/R_{substrate}$ at gate bias $V_g = +50$ V by that at $V_g = 0V$ (see Supporting Information Fig. S3 for two gating voltages). The excellent agreement between the transient reflectance spectrum from photo-exciting $WSe_2$ in the heterojunction and difference reflectance spectrum from n-type gate-doping unambiguously confirms electron transfer from $WSe_2$ to $MoS_2$ while the hole stays in the former. Generally, the injected electrons in the $MoS_2$ layer can affect its excitonic transitions in multiple ways,[22] including band filling (Pauli blocking), Coulombic screening, band renormalization, and the unique exciton-to-trion conversion in 2D monolayers[39]. Regardless of the detailed mechanism, electron transfer to $MoS_2$ is reflected by an overall bleach of the excitonic transitions in $MoS_2$. In the weak excitation regime probed here, we verify the initial transient reflection signal amplitude (at ~ 0.5 ps) shows a linear dependence on excitation density ($\leq 5 \times 10^{11}$ cm$^{-2}$, see Supporting Information, Fig. S4). Therefore, we take $\Delta R/R$ as proportional to electron population in the MoS$_2$ layer and probe the interfacial charge transfer dynamics.

Fig. 2C shows the early-time dynamics (red circles) probed at ~ 675 nm (bleach maximum of the MoS$_2$ A exciton) following photoexcitation of WSe$_2$ in the heterojunction. Also shown is the pump-probe laser cross-correlation (CC, grey dashed curve) with FWHM ~ 190 fs, as well as time-convoluted CC (black curve) which represents prompt rise in signal within experimental time resolution. The excellent agreement between experimental data (red triangles) and convoluted CC (black curve) establishes that electron transfer from photo-excited WSe$_2$ to MoS$_2$ occurs on the ultrafast time scale ≤ 40 fs (experimental time resolution, ~ 1/5 of the FWHM), in agreement with previous estimates from pump-probe spectroscopies[17,18] and absorption peak broadening.[40]

In the interfacial electron transfer process, the hole stays in the WSe$_2$ layer, as confirmed by a bleach of the WSe$_2$ excitonic transition position (see Supporting Information, Fig. S5). We use $\Delta R/R$ at the WSe$_2$ transition (W probe, ~ 754 nm) to follow hole population in the WSe$_2$ layer. Because of reduced dielectric screening in a 2D system, the hole left in WSe$_2$ and the electron transferred to MoS$_2$ should form an bound interlayer CT exciton, with binding energy > 100 meV.[15,16] We compare the electron (red circles) and hole (blue triangles) decay dynamics in Fig. 2D. The same dynamics, with a single-exponential fitted lifetime of $\tau_{CR}$ = 47 ± 9 ps, confirms that interlayer CT exciton decays through interfacial electron-hole charge recombination (CR). The CR process occurs with a ~3 times longer lifetime than that for the nonradiative exciton recombination in the WSe$_2$ monolayer, due to the spatially indirect nature of the CT exciton in the former.

We probe the dependence of interfacial charge transfer on the angular alignment (and thus parallel momentum mismatch) between the WSe$_2$ and the MoS$_2$ monolayer in the heterojunction using 12 samples with twist angle $\Delta\phi$ in the range of 0-30°. All the heterojunction samples probed here are encapsulated with h-BN capping layers, as shown by representative optical images for three h-BN capped WSe$_2$/MoS$_2$ heterojunction samples ($\Delta\phi$ = 7°, 17°, and 29°) in Fig. 3 A, B, C. The corresponding kinetics probing electron population in MoS$_2$ following photo-excitation of WSe$_2$ are shown in Figs. 3 D, E, F (see Supporting Information, Fig. S6 for results from other samples). Each kinetic profile is described by a prompt rise (≤ 40 fs) corresponding to ultrafast electron transfer from WSe$_2$ to MoS$_2$, and a much slower decay from interlayer charge recombination which can be fitted by a single exponential function. The lifetimes of interfacial CT and CR processes for all samples are shown in red semi-transparent area and with blue symbols, respectively, as a function Δϕ (bottom) or momentum change (top). Regardless of Δϕ, interfacial electron transfer in WSe$_2$/MoS$_2$ van der Waals heterojunctions all occurs on an ultrafast time scale within experimental time resolution (< 40 fs). This CT time is much shorter than the A exciton lifetime in the WSe$_2$ monolayer (16 - 68 ps in our samples), suggesting near-unity electron transfer efficiency at the van der Waals heterojunction. The charge recombination lifetime varies by almost three orders of magnitude from sample to sample, from ~ 47 ps to ~ 3 ns, but doesn't show any clear correlation with the twist angle. Within each sample, we find that the CR kinetics is insensitive to the position within the heterojunction region (Supporting Information, Fig. S7).

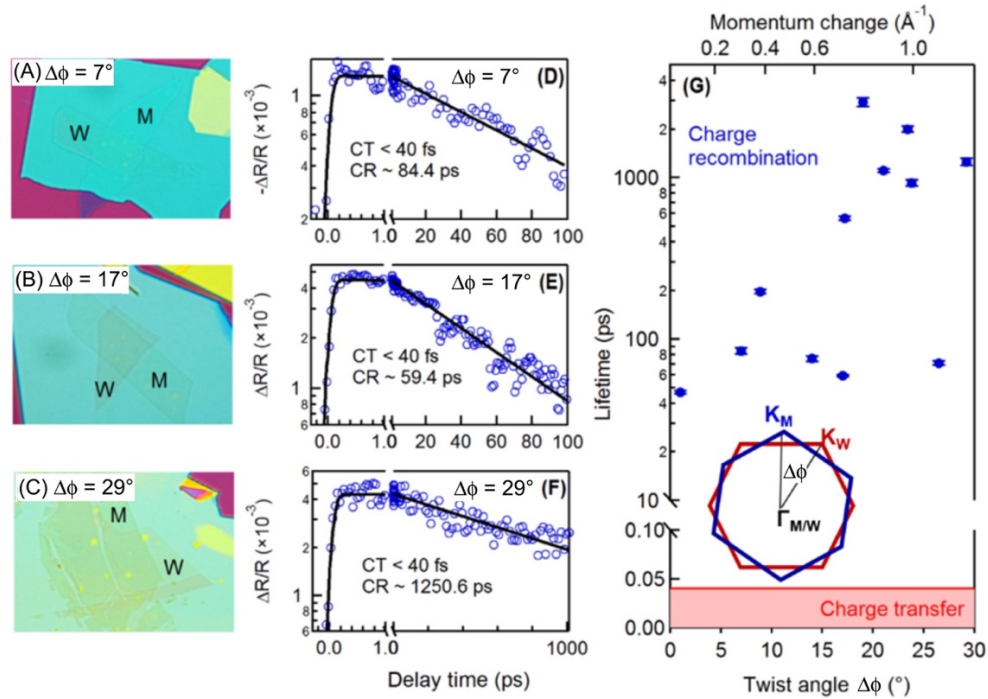

Fig. 3. Optical images (A-C) and corresponding transient reflection kinetics in the MoS$_2$ layer (D-F) for BN capped WSe$_2$/MoS$_2$ heterojunctions with three different twist angles: Δϕ = (D) 7°, (E) 17°, and (F) 29°. M: MoS$_2$; W: WSe$_2$. Each kinetic profile can be described by a single exponential rise (CT process) and a single exponential decay (CR process), convoluted with the laser CC function. Note the logarithm scale for vertical axis. (G) Interfacial charge transfer lifetime and charge recombination lifetime as a function of Δϕ (bottom axis) or momentum change (top axis) in WSe$_2$/MoS$_2$ heterojunctions. Note the logarithm scale for vertical axis above 10 ps lifetime. Inset: scheme illustrating the twist angle in momentum space.

The absence of correlation between CT/CR rates and Δϕ is surprising, giving the large momentum change. One possibility of momentum conservation comes from the bound nature of intra-layer and inter-layer excitons. The amount of momentum change shown in Fig. 3G is calculated based on the single particle picture. In reality, the reduced dielectric screening leads to tightly bound intra-layer and inter-layer excitons in the $WSe_2$ monolayer and at the $WSe_2/MoS_2$ heterojunction, respectively. Our estimates based on the uncertainty principle and spatial localization of the excitons give uncertainties in parallel momentum of 0.15 and 0.08 $Å^{-1}$ for the $WSe_2$ monolayer and $WSe_2/MoS_2$ heterojunction, respecrtilve;[15] these values are an order of magnitude too small to compensate momentum change accompanying the CT and CR processes.

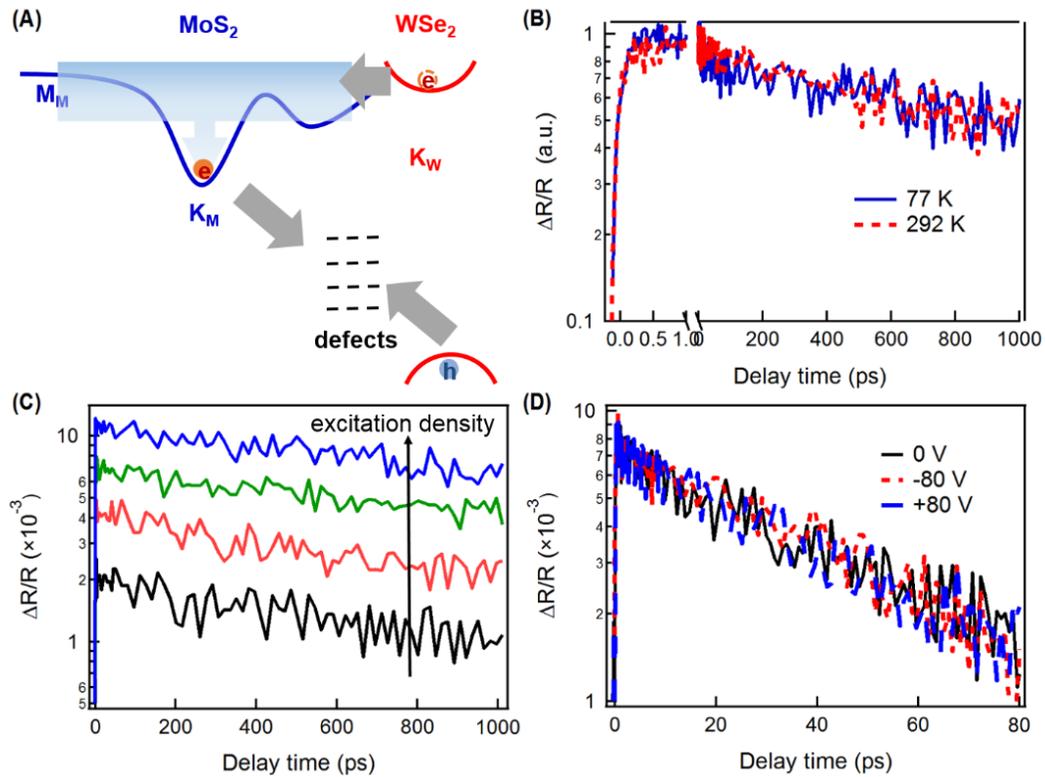

Fig. 4. (A) Schematic illustration of interfacial charge transfer process at the $WSe_2/MoS_2$ heterojunction. After photoexcitation, electrons transfer from $WSe_2$ to $MoS_2$ with excess energy. After intra-band electron relaxation, inter-layer electron-hole recombines through defect assisted nonradiative recombination. (B-C) Electron population kinetics in $MoS_2$ layer for a representative sample at different temperatures (B, 77 K and 292 K) and different excitation densities (C, from bottom to top 0.85, 1.7, 3.4, 5.1x$10^{11}$ $cm^{-2}$ ). (D) Charge recombination kinetics at different gating voltage which changes the doped electron density in the $MoS_2$ monolayer.

Since the initial CT rate is orders of magnitude faster than the exciton trapping process in the WSe$_2$ monolayer, we do not need to consider the role of defects. Instead, we attribute the $\Delta\phi$ independent CT rates to the excess electron energy at type II heterojunction, Fig. 4A. The conduction band offset between WSe$_2$ and MoS$_2$ is large (0.76 eV).[26] The excess energy means that electron transfer process can sample a broader range of K$_\parallel$ space above CBM,[32] including the M point which intersects the K valley in WSe$_2$ at $\Delta\phi = 30°$. A recent ultrafast infrared spectroscopy study also suggests the presence of hot interlayer excitons at the heterojunction interface.[41] The large excess electronic energy in the interfacial CT process also accounts for the apparent temperature independence in the initial ultrafast CT rates (Fig. 4B). The electron transfer event is followed by hot electron relaxation within the MoS$_2$ conduction band. The lowest energy direct gap nature at K valley in MoS$_2$ is well retained due to weak interlayer coupling.[16,26] After charge transfer, the electron is expected to relax to the K valley CBM in MoS$_2$ with the hole in the K valley CBM in WSe2, forming an interlayer CT exciton.

Unlike the electron transfer process, the recombination between the electron at the MoS$_2$ CBM and the hole at the WSe$_2$ VBM occurs with well-defined momentum vectors and has to be accompanied by momentum change if they recombine directly. An electron-hole pair in a semiconductor or a semiconductor interface recombines via three likely channels: (i) radiatively, (ii) nonradiatively through Auger recombination, or (iii) nonradiatively via defect-assisted recombination. The radiative mechanism is not important here as CT exciton emission is not observed in our heterojunction samples. The Auger mechanism refers to the nonradiative recombination of the CT exciton with both energy and momentum transferred to another CT exciton or free carrier. Fig. 4C shows CR kinetics of a representative heterojunction sample as a function of excitation density (see Supporting Information, Fig. S6, for other samples). The independence of CR kinetics on excitation density precludes exciton-exciton Auger recombination in the excitation density range probed here ($\leq 5 * 10^{11}$ cm$^{-12}$). In contrast, we observe faster recombination for intra-layer excitons in WSe$_2$ monolayer in the same excitation density range (see Supporting Information, Fig. S4). We also rule out exciton-carrier based Auger scattering, as the CR dynamics is found to be independent of gate-doping of MoS$_2$, with either electron doping at $V_g = + 80$ V or hole doping at $V_g = - 80$V, corresponding to electron and hole doping, respectively, of $6.2 \times 10^{12}$ cm$^{-2}$, Fig. 4D.

These results suggest that interlayer electron-hole recombination at the $MoS_2/WSe_2$ heterojunction is dominated by defect-assisted nonradiative electron-hole recombination, also known as the Shockley-Read-Hall mechanism.[42,43] Previous studies on TMDC monolayers have shown the detrimental role of defects on carrier dynamics and photoluminance efficiencies.[35,37,44] The tightly localized nature of trapped carriers in defects can satisfy the momentum conservation for both carrier trapping and recombination, with energy likely released through multi-phonon emission process. The properties of defect, including energetics, distributions, and densities, can vary from sample to sample from the transfer-stacking process, leading to large variations in inter-layer CT exciton recombination rates. Thus, in order to probe the intrinsic physics of momentum conservation in radiative CT exciton emission, one must not only control the inter-layer angular alignment within the range defined by the light cones,[9,21] but also minimize trap densities at the TMDC heterojunction. The nature of these defects, how they exactly assist interlayer exciton recombination and how we may control them remain challenging questions that deserve answers in future experiments.

In summary, our findings suggest the following mechanisms for photo-induced charge separation and recombination at 2D TMDC heterojunctions with type II band alignment. Intra-layer excitons from photoexcitation are dissociated with ultrafast rates and near unity yield by resonant interfacial charge transfer, with injected charges carrying excess kinetic energy. After cooling to the band edges (K valleys), the interlayer CT exciton recombines nonradiatively through defect assisted recombination. The initial interfacial charge separation and the subsequent interfacial charge recombination circumvent momentum mismatch via excess electronic energy and via defects, respectively.

**Methods**

Sample preparation.

We prepared $WSe_2/MoS_2$ heterojunctions using the polydimethylsiloxane (PDMS) transfer technique for grated sample[28] and the hexagonal boron nitride (h-BN) pickup/transfer technique[29] for other samples. In the first method, monolayer $WSe_2$ flakes were mechanically exfoliated onto 285nm $SiO_2/Si$ chips and monolayer $MoS_2$ onto PDMS stamps using mechanical exfoliation with Scotch tape. Then, the MoS$_2$ flake on the PDMS stamp was inverted and aligned onto the target WSe$_2$ monolayer on the SiO$_2$ substrate by micro-manipulator and held in contact for 5 minutes at 40 °C in order to transfer the flake to the substrate. For electric gating measurements, we fabricated electric contacts (Al / Cr / Au) on the MoS$_2$ monolayer region using e-beam lithography (no contact on WSe$_2$) and the gate voltage was applied by a Keithley 2400 power supply/analyzer. The successful gating on MoS$_2$ layer can be confirmed by PL spectra (Supporting Information, Fig. S3). In the second method, monolayers of MoS$_2$ and WSe$_2$ on the SiO$_2$/Si substrate were obtained by mechanical exfoliation and identified under an optical microscope. A relatively thick (about 20-50 nm) h-BN layer was first exfoliated and picked up by a small piece of PDMS with a thin layer of poly-propylene carbonate (PPC) on it. Then the h-BN was used to pick up the MoS$_2$ monolayer flake and then stamped onto the target WSe$_2$ monolayer. The substrate was kept at 120°C for 5 minutes before the PDMS was lifted up. As the PPC has melted completely at this temperature, the BN-capped TMDC heterojunction stayed on the substrate after the PDMS was lifted. The sample was then rinsed in acetone for 5 minutes to remove any PPC residue on top of the h-BN, resulting in a van der Waals stack of BN/MoS$_2$/WSe$_2$ on the SiO$_2$/Si substrate. All as-prepared samples were annealed at 250°C for 4 hours under high vacuum (10$^{-8}$ mbar).

Optical Characterization. The photoluminescence and reflectance measurements were performed on a home-build inverted microscopy system at room temperature. In photoluminescence measurements, the excitation light (532 nm) was focused with a 100X objective to < 1μm on different regions of asamples and the emission was collected by the same objective, diffracted by grating and detected on a CCD camera (PyLon400 for visible and PyLon IR for near-IR with IsoPlane 160 spectrograph, Princeton Instruments). In the reflectance measurements, the broadband white light from a Halogen Lamp was focused, reshaped with a pinhole, and then focused with a 100X objective to ~ 2 μm on different regions of samples. The reflected white light was collected by the same objective and detected with a CCD camera (PyLon400, Princeton Instruments). The reflectance (R) was calculated as the ratio between reflected light from sample and that from the substrate ($R_{sample}$/$R_{substrate}$).

SHG measurements. The SHG measurements were performed on a home-build inverted microscopy system. A linear polarized femtosecond laser light (800 nm, 80 MHz, 100 fs, 100 uW, Coherent Mira) was focused with a 100X objective on MoS$_2$ or WSe$_2$ monolayer region. The reflected 400 nm light was collected by the same objective and detected by a PMT detector (Hamamatsu R4220P) after passing through a linear polarizer and a 400 nm narrow band filter. We fix the polarization direction of excitation light and collection light and rotate the heterojunction sample. The absolute rotational angle is in arbitrary unit. We fit the angular dependent SHG intensity $I$ using equation $I = I_{\max}\cos^2[3(\theta + \varphi)]$. Because SHG intensity measurement is not sensitive to phase, it leads to a six-fold symmetry pattern and $\varphi$ is between 0 and 30 degree. Therefore, the extracted twist angle $\Delta\phi = |\varphi_W - \varphi_M|$ for all heterojunctions varies between 0° and 30°.

Micro pump-probe measurements. The micro pump-probe studies were performed on a home-build inverted microscope system. The wavelength tunable femtoseconbd excitation light was generated from a femtosecond laser system (Coherent RegA, 800 nm, 250 KHz, 100 fs, 1.2 W) pumped OPA (Coherent OPA 9450). The probe light was a white light continuum generated by focusing a small portion of the fundamental light onto a Sapphire window (1mm). Both beams were collinearly aligned and focused with a 100X objective onto different regions of each sample. The pump beam was adjusted to be ~ 2 μm and probe beam ~ 0.8 μm, to ensure homogeneous excitation and to remove effects due to carrier in-plane diffusion. The reflected light was collected and detected with a PMT detector after passing through a monochromator. The excitation light was chopped by an optical chopper at 5 KHz and the transient reflection signal (ΔR/R) due to excitation was measured with a lock-in-amplifier (Stanford Research Systems). The time delay between the pump and the probe beam was controlled by a motorized delay stage. The pump fluence was kept below 3.5 μJ cm$^{-2}$ in all experiments, corresponding to excitation density 5*10$^{11}$ cm$^{-2}$ and the probe fluence was about one order of magnitude lower than that of the pump.

**Acknowledgements.** HZ and JW carried out all spectroscopic measurements: HZ was a fellow of the MRSEC program through the Center for Precision Assembly of Superstratic and Superatomic Solids supported by the National Science Foundation (NSF) grant DMR-1420634 and JW was supported by the NSF grant DMR 1608437. Sample preparation was supported by the MRSEC supported by NSF DMR-1420634. XYZ acknowledges support by the Samsung Global Research Initiative for the development of the pump-probe microscope setup.

SUPPORTING INFORMATION

# Interfacial Charge Transfer Circumventing Momentum Mismatch at 2D van der Waals Heterojunctions

Haiming Zhu[1,2,*], Jue Wang[1,*], Zizhou Gong[3], Young Duck Kim[4], Martin Gustafsson[1], James Hone[4], Xiaoyang Zhu[1,†]

[1.] Department of Chemistry, Columbia University, New York, NY 10027, USA

[2.] Department of Chemistry, Zhejiang University, Hanzhou, Zhejiang, 310027, China

[3.] Department of Physics, Columbia University, New York, NY 10027, USA

[4.] Department of Mechanical Engineering, Columbia University, New York, NY 10027, USA

[*] These authors contributed equally.
[†] To whom correspondence should be addressed. E-mail: xyzhu@columbia.edu

1. Optical image, SHG intensity as a function of rotational angle and PL spectra of all BN capped WSe$_2$/MoS$_2$ samples

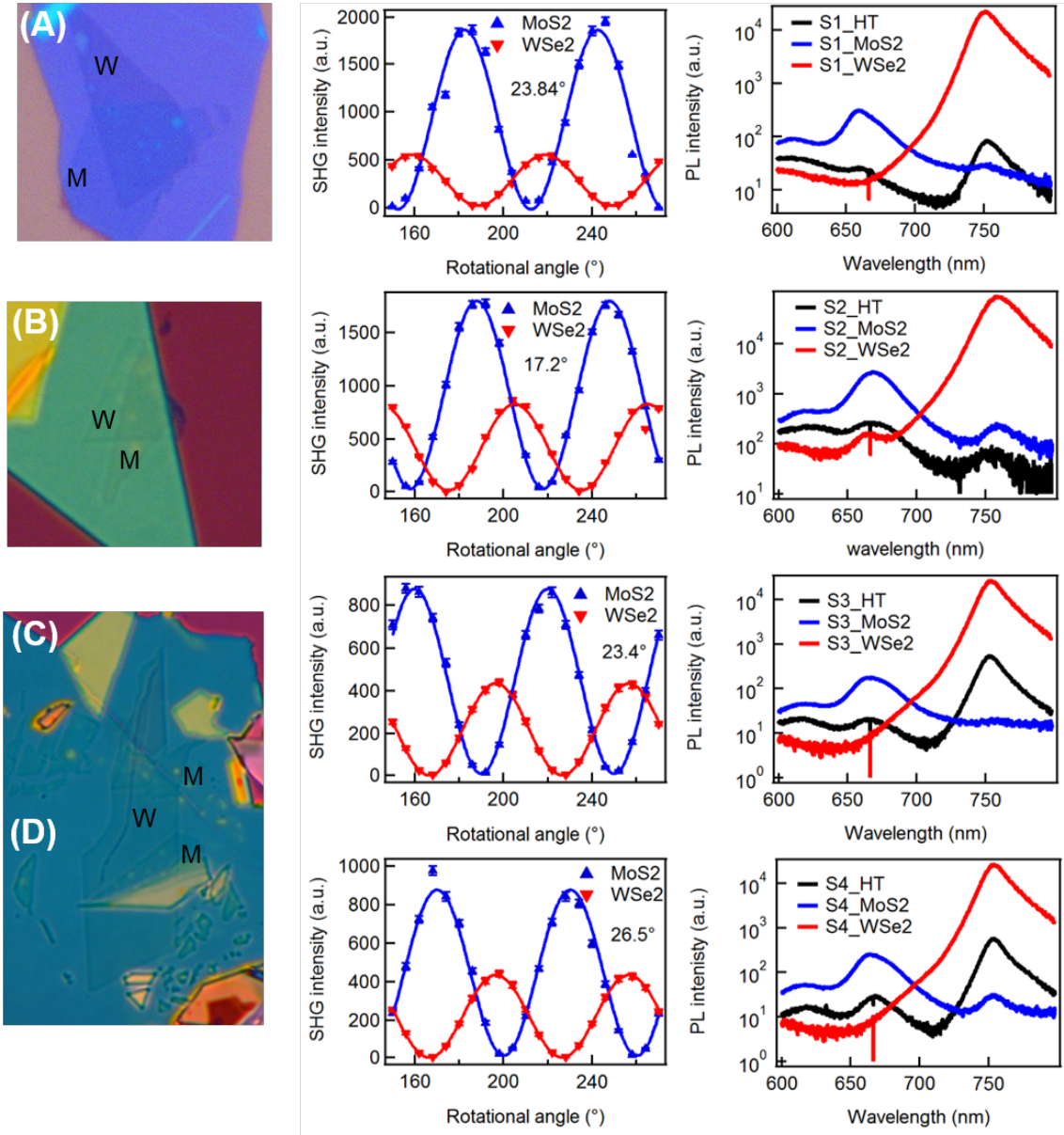

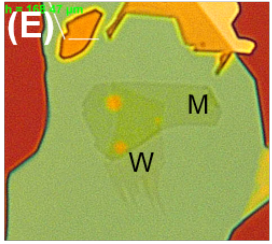
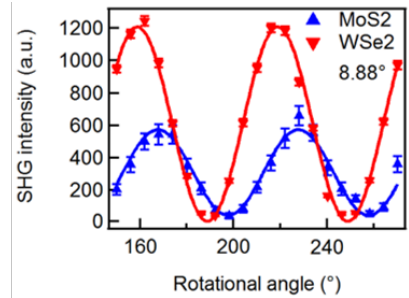
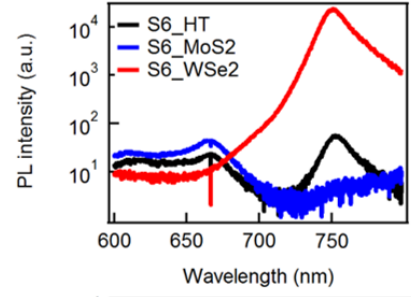

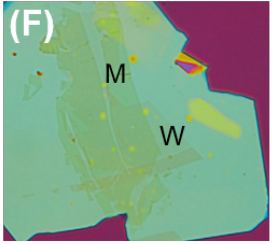
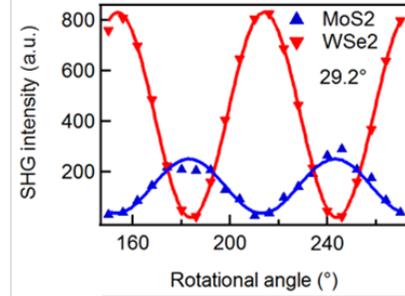
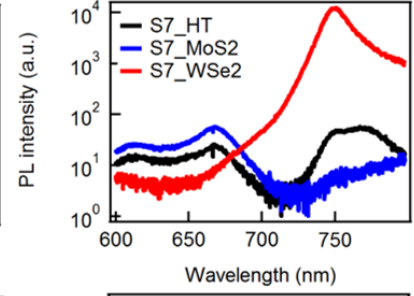

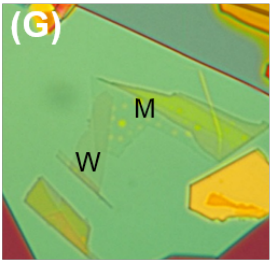
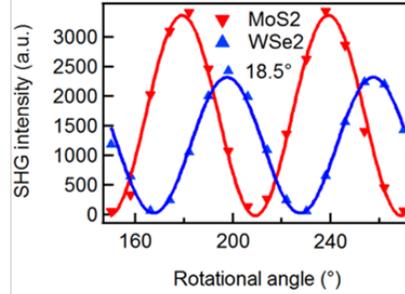
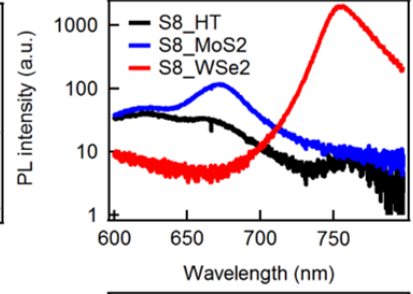

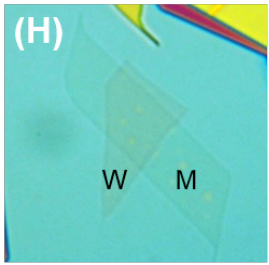
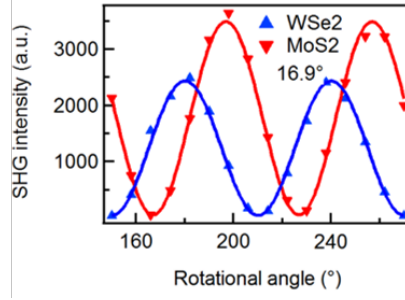
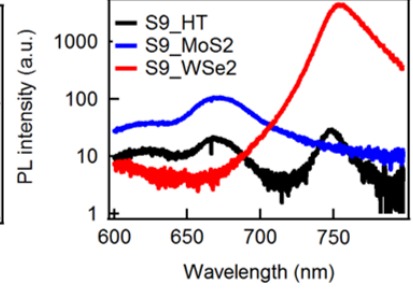

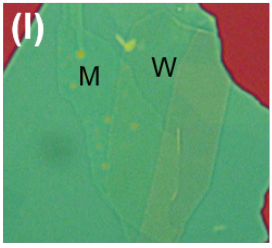
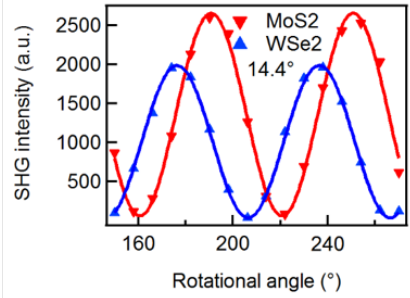
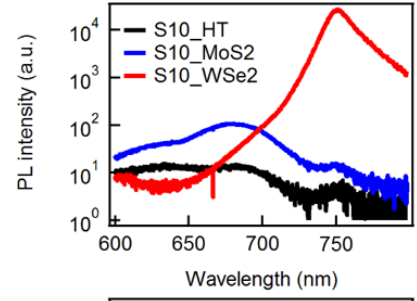
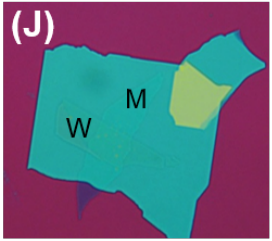
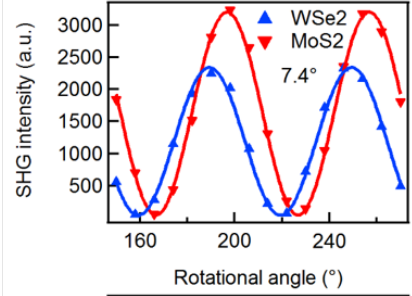
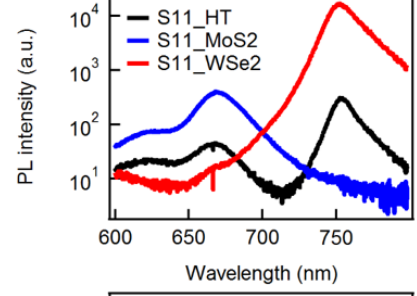
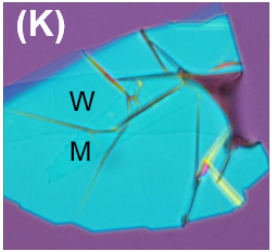
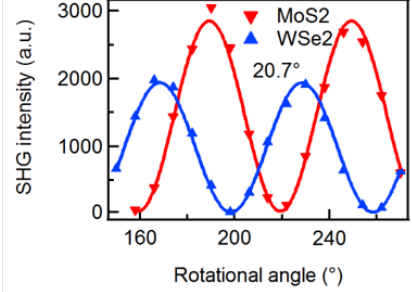
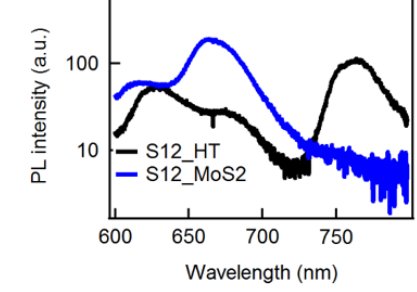

Fig. S1. (A-K) Optical image, SHG intensity as a function of rotational angle and PL spectra of all BN capped WSe$_2$/MoS$_2$ samples. The twist angles deduced are noted in SHG intensity plot.

2. Transient reflectance spectrum of WSe2 monolayer under 710 nm excitation

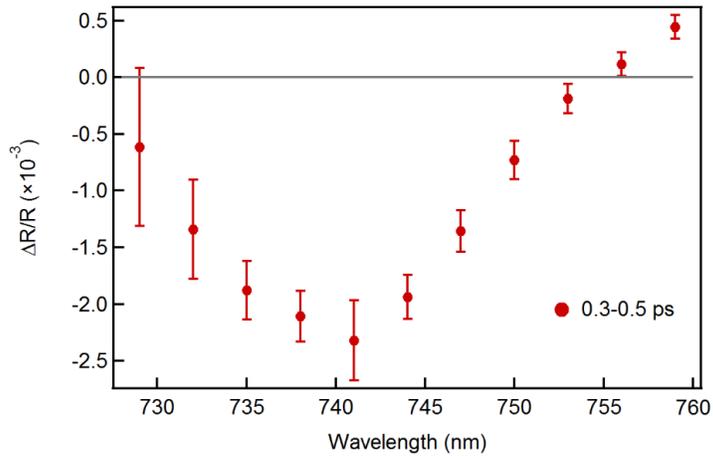

Fig. S2. Transient reflectance spectrum of WSe$_2$ monolayer under 710 nm excitation showing a bleach at ~ 740 nm.

3. Gating bias dependent PL spectra and reflectance spectra of $MoS_2$ layer in $WSe_2/MoS_2$ monolayer

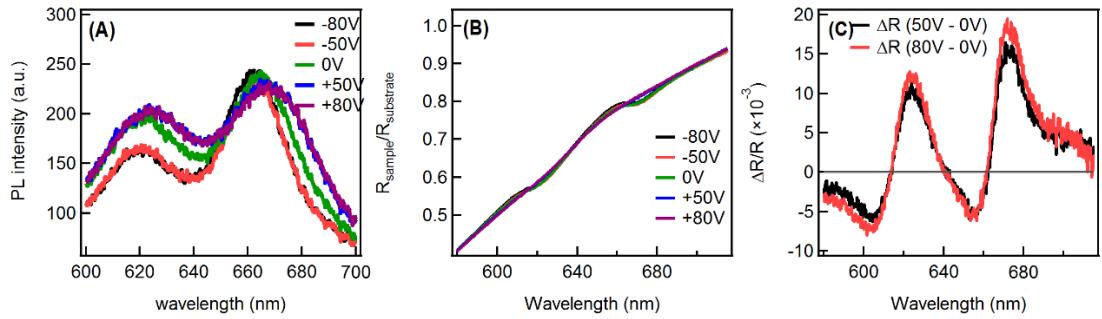

Fig. S3. Gating bias dependent PL spectra and reflectance spectra of $MoS_2$ layer in $WSe_2/MoS_2$ monolayer

4. Excitation density dependent kinetics of WSe2 monolayer and WSe2/MoS2 heterostructure of a representative sample

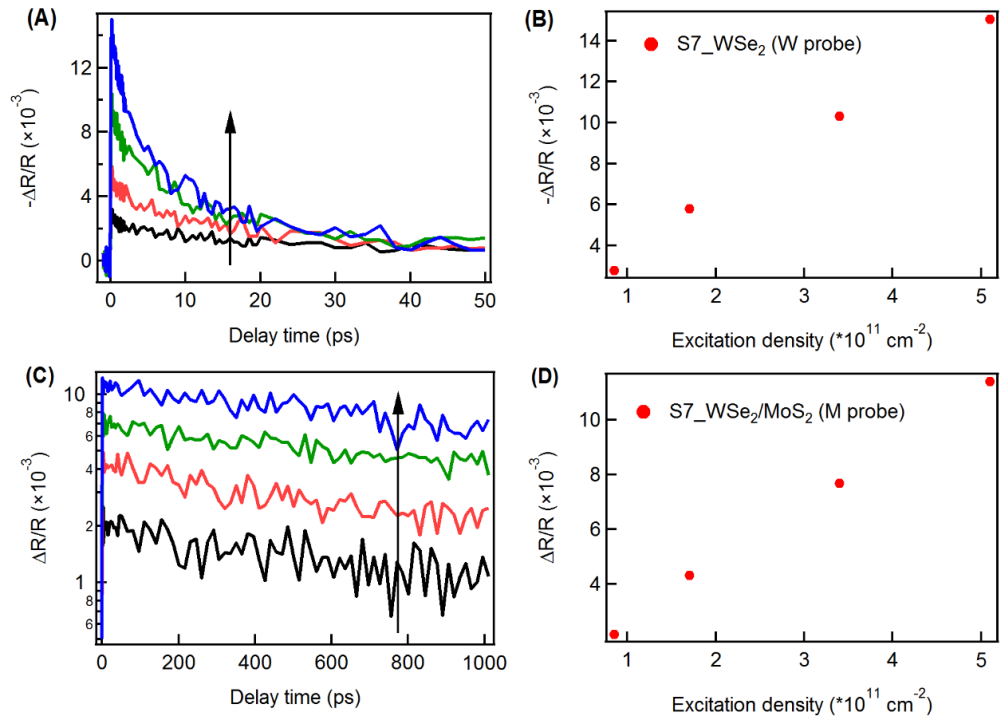

Fig. S4. Excitation density dependent transient reflectance kinetics of WSe2 monolayer (A, W probe) and WSe2/MoS2 heterostructure (C, M probe) of a representative sample. The excitation density is (0.85, 1.7, 3.4, 5.1) * $10^{11}$ $cm^{-2}$ from bottom to top. (B, D) Initial transient reflectance signal as a function of excitation density, showing linear relationship.

5. Early time transient reflectance spectra of WSe$_2$/MoS$_2$ and BN capped WSe$_2$/MoS$_2$ at WSe$_2$ probing wavelength

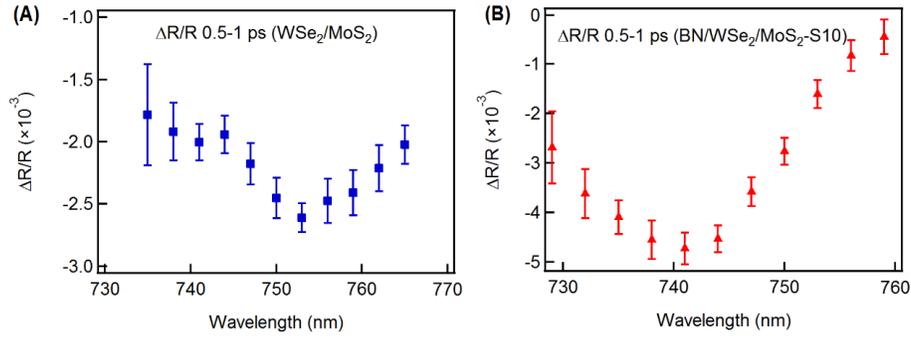

Fig. S5. Early time transient reflectance spectra of WSe$_2$/MoS$_2$ (A) and BN capped WSe$_2$/MoS$_2$ (B) at WSe$_2$ probing wavelength.

6. Transient reflectance kinetics in all BN capped WSe$_2$/MoS$_2$ samples

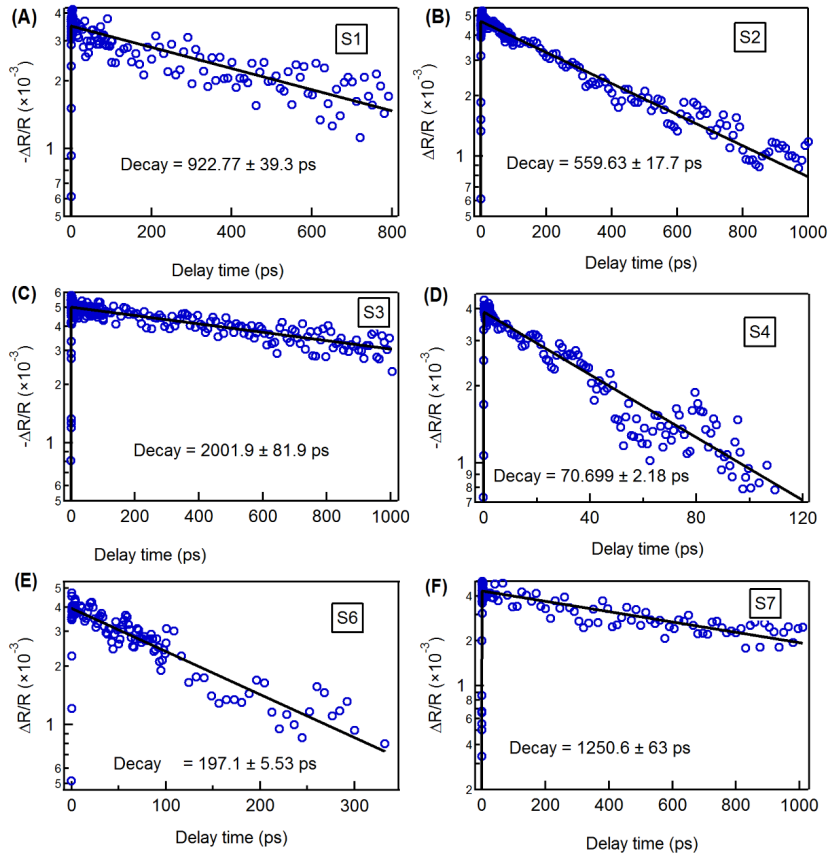

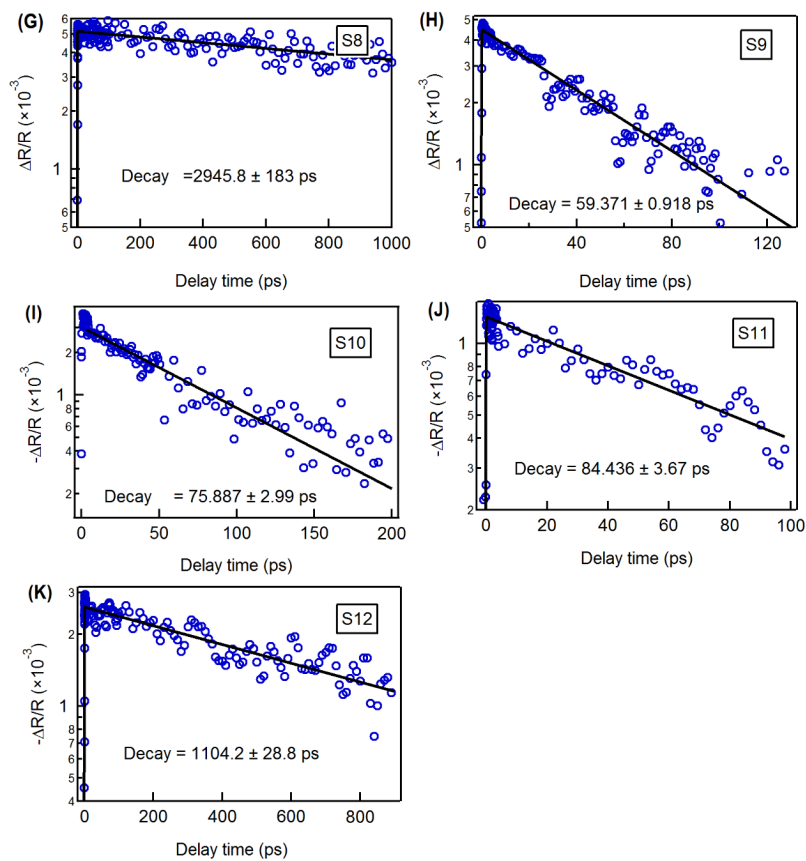

Fig. S6. Transient reflectance kinetics in all BN capped WSe$_2$/MoS$_2$ heterostructure samples (M probe) showing ultrafast electron injection and slow charge recombination.

7. Transient reflectance kinetics at different spots for two representative BN capped WSe$_2$/MoS$_2$ samples

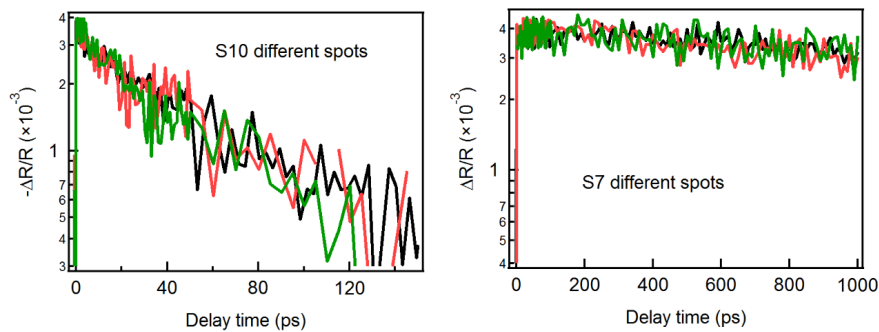

Fig. S7. Transient reflectance kinetics at different spots for two representative BN capped WSe$_2$/MoS$_2$ samples showing no dependence on positions.